# Using Optimization Algorithms for Control of Multiple Output DC-DC Converters

Masoud Safarishaal*, Mohammad Sarvi[1], **

Electrical Engineering Department

Imam Khomeini International University

Qazvin, Iran

*masoud.safarishaal66@gmail.com, **sarvi@eng.ikiu.ac.ir,

**Abstract:** The weighted voltage mode control represents a method for control of multiple outputs DC-DC converters. Accordingly, the weighted control redistributes the error among the outputs of these converters, and the regulation error can be reduced by adjusting the weighting factors. But the problem is that most designs are performed on the trial-and-error basis, and the results were rather inconsistent. Also, in conventional mathematical approaches, this factor is designed for converters by given parameters. In this paper, three optimization algorithms namely Imperialist Competitive Algorithm (ICA), Particle Swarm Optimization (PSO) and Ant Colony Optimization (ACO) are proposed for a quick and accurate estimation of the optimal weighting factors and improve the amount of regulation on outputs of multiple outputs forward DC-DC converters. Furthermore, Fuzzy Logic Controller (FLC) is utilized to minimize the total steady-state error and improve transient characteristics by controlling the duty cycle of the PWM controller. Simulations have been performed in several cases and results show that the proposed method improves the outputs cross regulations in multiple outputs forward DC-DC converters significantly. ICA based weighting factor estimator has higher speed and accuracy in comparison with two other presented algorithms (ACO and PSO) and so is more effective in comparison with them.

**Keywords:** Multiple Outputs Forward DC-DC Converters; Imperialist Competitive Algorithm; Particle Swarm Optimization; Ant Colony Optimization; Weighting Factor Method.

---

[1] Corresponding author: Mohammad Sarvi, Email: sarvi@eng.ikiu.ac.ir, Tel: +98-28-33901294, Fax: +98-28-33780073.

## 1. Introduction

Multiple output DC-DC converters are more efficient in comparison with using several separate single output power supplies [1]. These converters are employed in a variety of applications, including power supplies for personal computers, office equipment, spacecraft power systems, laptop computers, and telecommunications equipment, as well as dc motor drives [2]. One standard topology of these converters is multiple outputs forward DC-DC converters. In the past years, the interest in direct current to direct current converters have increased because of their application in renewable energy systems. However, one of its major drawbacks of multiple output DC-DC converters in common is the poor regulation. Typically finding the optimal duty cycle combination is not an easy task. A variety of methods have been presented in several published works of literature which discuss the various issues of cross regulations [3-5].

From a control point view, multiple output converters are controlled in one of the most important of their outputs (as namely Master), and the rest of the outputs (as namely Slave) intensively depend on the regulation of the main (Master) output [6]. All the slave modules have the same controller structure and receive the reference voltage from the master module. It means the main voltage regulation is made from the master voltage. This subject is a limitation to control all output voltages when input voltage or load fluctuates.

Another typical method for regulating a multiple outputs converter which herein referred to as weighted voltage mode control is controlling of all output simultaneously and feedback the sum of the weighted outputs. The weighted control redistributes the error among the outputs of multiple outputs forward DC-DC converters, and the regulation error can be adjusted by varying the weighting factors. Simultaneous improvements of the transient and dc performance of the cross-regulated outputs can be obtained by using weighted voltage mode control [7-8].

Although the weighted voltage mode control has been used in some literature, most designs are performed on the trial-and-error basis, and the results were rather inconsistent [9-10]. If the weighting factors and the compensator are properly designed, the dc cross-regulation and dynamic characteristics of the outputs can be significantly improved. Generally, the tightness of cross-regulation is dependent on a number of the circuit parasitic and for a given power stage the output voltages are functions of input voltage, load currents and weighting factor [11]. In the traditional mathematical based methods of the weighting factor determination, this factor is just designed for the converter by given parameters. However, when the major parasitic on dc output voltage has been changed, it is very difficult to calculate that factor again. This paper proposed general, fast, and accurate methods to the determination of the weighting factors to meet the desired dc regulation. Accordingly, use of optimization algorithms has been proposed to improve the amount of regulation in all outputs by determination of best weighting factors simultaneously by finding the optimal duty cycle combination. In this paper, as an alternative to the conventional mathematical approaches, three evolutionary algorithms namely Imperialist Competitive Algorithm (ICA), Particle Swarm Optimization (PSO) and Ant Colony Optimization (ACO) have been utilized to get optimum weighting factors. Recently, ICA has been successfully employed in many engineering applications [11-12]. The PSO is a very good population-based, bio-inspired optimization method which used in the different optimization problem [12-14]. Moreover, ACO has been successfully applied to many difficult problems [15-17]. In the proposed method, the algorithm finds the best weighting factors itself, accurately and quickly. The results show efficiency and capabilities of the proposed algorithm in finding the optimum design. Such a control topology is applied not only to the forward converter but also to any isolated or non-isolated converter or any inverter.

In addition, Fuzzy Logic Controller (FLC) is utilized to minimize the total steady-state error and improve transient characteristics by controlling the duty cycle. FLC has received increasing attention from researchers for converter control, as it provides better responses than other conventional controllers [18-19].

The rest sections of the paper are organized as follows: Section II gives the proposed control method description. The evolutionary algorithm-based weighting factor is explained in Section III. Section IV presents the results and discussions. Finally, conclusions are given in Section V.

## 2. The Proposed Control Method Description

In this paper, a forward topology with three outputs and weighted error voltage mode control approach is considered. All three outputs voltages are sensed and compared with reference voltages, and then the resulting error multiplied by three weight factors. By using optimization algorithms (ICA, PSO, and ACO) the most appropriate parameter of these factors can be calculated for the considered system during operation. They are computational methods that optimize a problem by iteratively trying to improve a candidate solution with regard to a given measure of quality. In this way, there has been much more improved performance with the proposed tuned controller compared with having fixed factors in term of good regulation. So, for setting weighting factors, proposed algorithms are applied. The optimization algorithms will repeat the illustrated steps until they achieve the optimal solution for the minimization of the objective function. These algorithms are coded in MATLAB and are helping the factor to guide the exploration and increase the run with a computer. In all of the examples, control in the exploitation the population is equal to 50. Also, a fuzzy logic controller is utilized to minimize the total steady-state error and improve transient characteristics. Fig.1 shows the system block diagram as well as the proposed method.

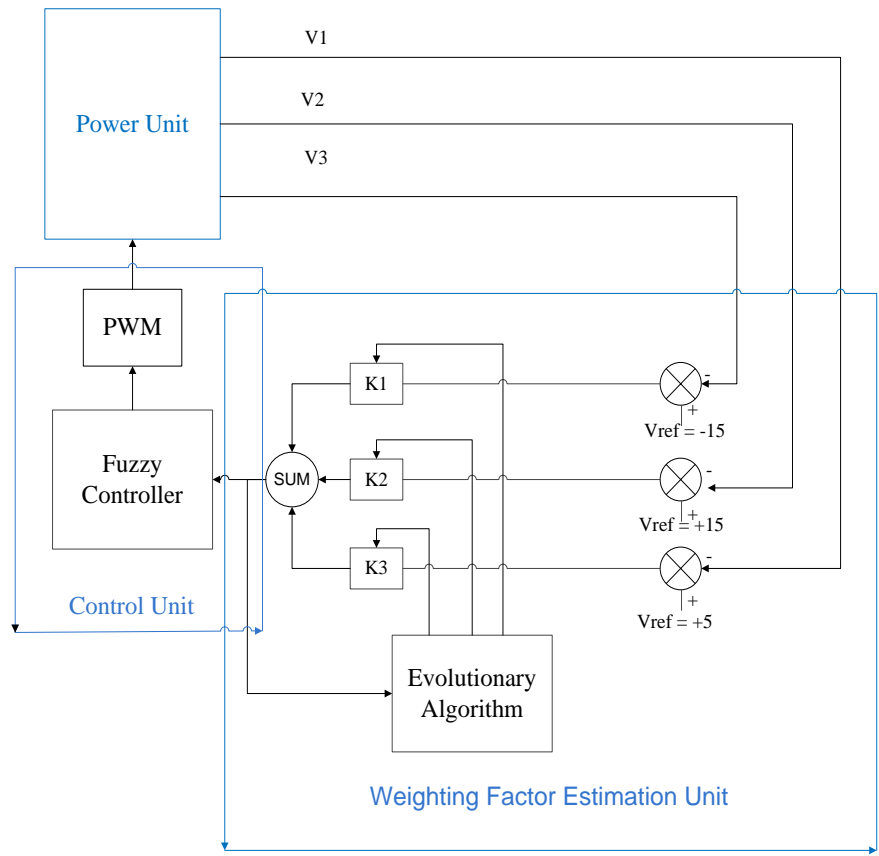

**Fig.1.** The proposed control strategy.

The topology of the proposed system consists of three main parts: power unit, control unit, and weighting factor estimation unit. Power unit and control unit are explained in detail in the following. Also, the weighting factor estimation unit is presented in section 3.

### 2.1. Power Unit

The power stage block is consisting of switching transistors (consist of two switching MOSFETs), high-frequency transformer, output rectifiers, and output filters. In this paper a converter with three outputs (5V/50W, 15V/45W and -15V/15W) under operation of input voltage range from 18V to 40V dc and a switching frequency of 50 kHz are considered. The transient characteristics of the cross-regulated outputs can be improved by coupling the output filter inductors. Fig.2 shows these multiple outputs forward DC-DC converters. This circuit uses a single DC input (nominally 28 Volt) and converts to three simultaneous output voltages.

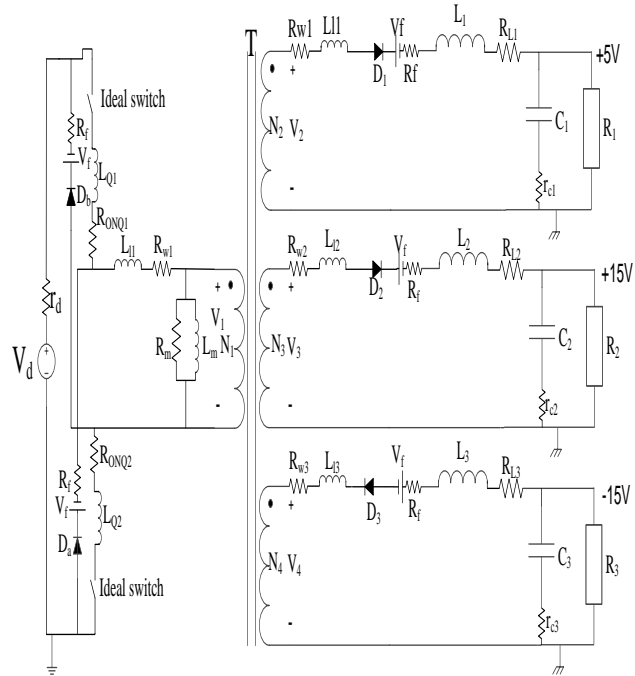

**Fig.2.** Multiple outputs forward DC-DC converter.

In Fig 2, $r_d$ is the source resistance; $R_{w1}$ and $R_{wi}$ are the primary and secondary winding resistances, respectively. $R_f$ is the ON state resistance of $D_1$. $R_{ONQ1}$ and $R_{ONQ2}$ is the ON state resistance of switching MOSFETS, respectively. $L_{11}$ and $L_{22}$ are the primary and secondary leakage inductances, respectively. $L_{Q1}$ and $L_{Q2}$ are the MOSFETS leakage inductances, respectively. $r_{ci}$ is the Equivalent Series Resistor (ESR) of the output capacitor [20].

### 2.2. Control Unit

Control unit is consisting of a Fuzzy Logic Controller and Pulse-Width Modulation (PWM) generator. The power switches are controlled by PWM technique, and the pulses are generated by the application of fuzzy logic.

### 2.2.1. Fuzzy Logic Controller

Fuzzy control has been applied to control a dc-dc forward converter. Fuzzy controllers are well suited to nonlinear time-variant systems and do not need an exact mathematical model for the system being controlled. They are usually designed based on expert knowledge of the converters. [20]

The inputs to the FLC are error signal and difference of error signal. The output is the duty ratio of the switching signal.

$$E(t) = V_{ref} - V_o(t) \tag{1}$$

$$dE = E(t) - E(t-1) \tag{2}$$

$$dD = D(t) - D(t-1) \tag{3}$$

Where $V_{ref}$ is reference voltage, $V_o(t)$ is output voltage at $t^{th}$ instant, $E(t)$ and $E(t-1)$ are error signal, at $t^{th}$ and $(t-1)^{th}$ instant, respectively. $D(t)$ and $D(t-1)$ are duty cycle at $t^{th}$ and $(t-1)^{th}$ instant, respectively. *dE* and *dD* are change in error and duty cycle, respectively.

Mamdani type controller is chosen for this application. To obtain the control decision, the max-min inference method is used. Also, the center of gravity method of defuzzification is applied to obtain the crisp result from the linguistic values obtained according to the rule base. The basic rule of this type of controller is: IF *E* is A and *dE* is B THEN *d (t)* is C.

Where A and B are fuzzy subsets, C is a fuzzy singleton. The membership functions of the inputs error, change in error and output are triangular as shown in Fig.3. Inference system 7×7 rule base used for the design of the fuzzy controller is shown in Table 1 for illustration purposes.

Table 1. Rule base used in the fuzzy controller.

| E\dE | NB | NM | NS | ZE | PS | PM | PB |
|---|---|---|---|---|---|---|---|
| **NB** | NB | NB | NB | NB | NM | NS | ZE |
| **NM** | NB | NB | NB | NM | NS | ZE | PS |
| **NS** | NB | NB | NM | NS | ZE | PS | PM |
| **ZE** | NB | NM | NS | ZE | PS | PM | PB |
| **PS** | NM | NS | ZE | PS | PM | PB | PB |
| **PM** | NS | ZE | PS | PM | PB | PB | PB |
| **PB** | ZE | PS | PM | PB | PB | PB | PB |

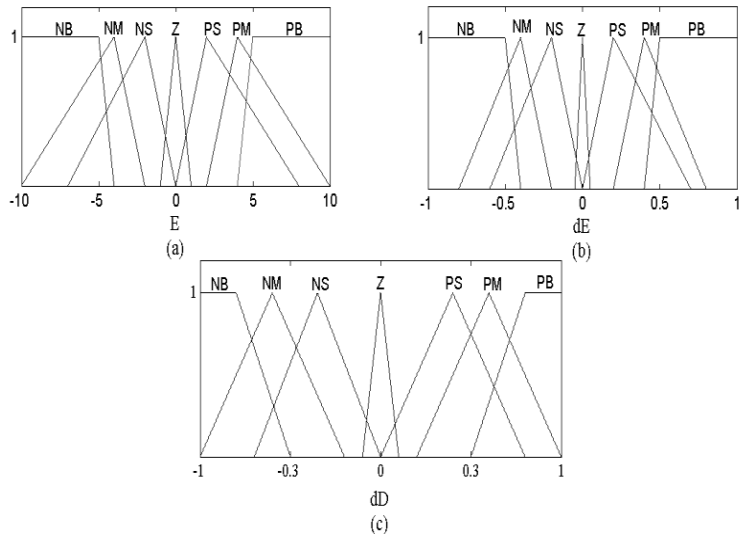

**Fig.3.** Membership functions of (a) input *E*; (b) input *CE*; (c) output *dD*.

In table 1, *N* stands for negative, *P* stands for positive, and *Z* represents zero. *B* means big, *M* means medium and *S* stands for small. For example, *NB* means negative big.

### 2.2.2. Pulse Width Modulation (PWM)

PWM generates the pulse signals for the converter based on the desired duty cycle when fuzzy logic computes the desired duty cycle of the converter using total steady-state error. The duty cycle is the control input, this input is a logic signal which controls the switching action of the power stage and hence the output voltage.

### 3. Evolutionary Algorithm Based Weighting Factor Estimator

For a multiple output DC-DC converter with weighted voltage mode control, each output is not only related to the circuit parameters, but also to the weighting factor ($K_i$) [21]. In this paper, the proposed optimization algorithms estimate the optimal weighting factor corresponding to all three output voltages, in order to improve the amount of regulation on outputs. Three evolutionary algorithms are ICA, PSO and ACO. These optimization algorithms will repeat the illustrated steps until it finds the optimal solution for the minimization of the fitness function.

The fitness function for the algorithms is considered corresponding to the output errors. It is expressed in the sum of the absolute terms of the relative errors as following:

$$FitnessFunction = |e_{+15}| + |e_{+5}| + |e_{-15}| \tag{4}$$

Where, $e_{+5}$ is the +5 v output error, $e_{+15}$ is the +15 v output error and $e_{-15}$ is the -15 v output error.

In artificial intelligence, an evolutionary algorithm is a subset of evolutionary computation, a generic population-based metaheuristic optimization algorithm.

### 3.1. Imperialist Competitive Algorithm (ICA)

The ICA algorithm is colonial competition inspired by the idea of human socio-political evolution. In the original algorithm, the number of colonial countries together with their colonial countries seeks to naturally find the general optimal point to solve efficiently the optimization problem. In this paper, the number of established colonies is assumed to be Nc=50, and the number of initial imperialists is assumed to be Np=20 also the stopping criteria is defined as reaching to the maximum number of iterations (Max Iteration=100) and when no more than one imperialist exists in the search space. Fig.4 typically shows the flowchart of the ICA.

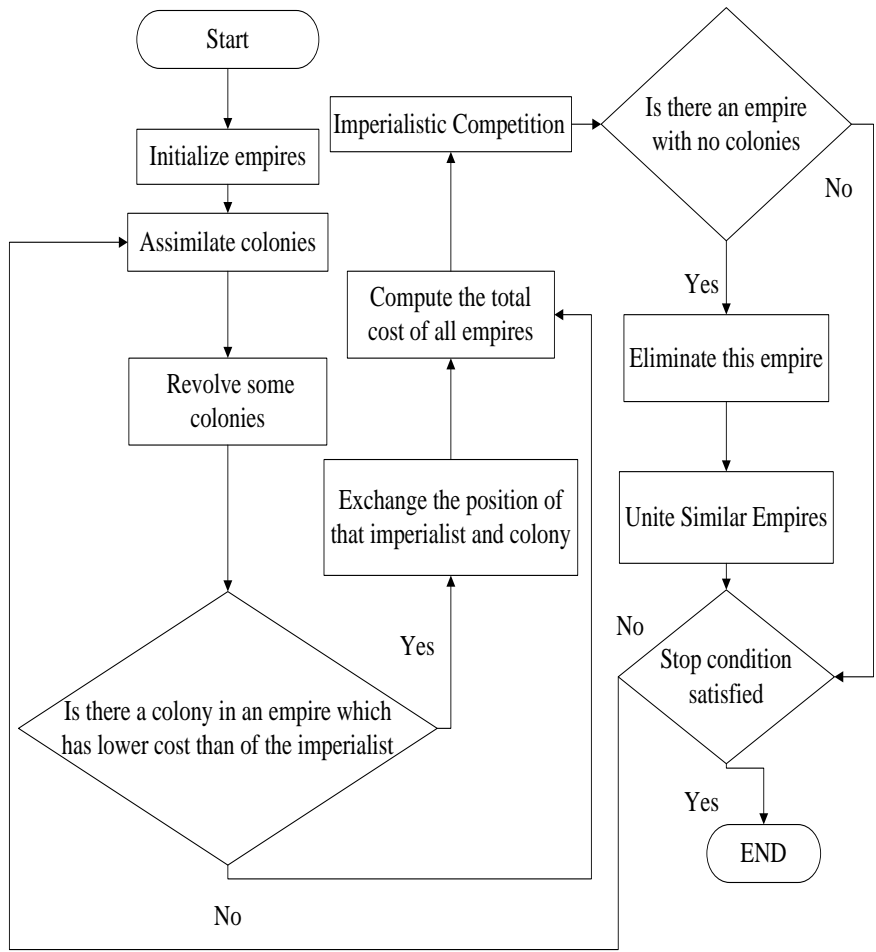

**Fig.4.** Flowchart of ICA.

### *3.2. Particle Swarm Optimization (PSO)*

PSO is a global optimization method that can be used to deal with problems whose answer is a point or surface in the next n space. In such a space, hypotheses are made, and an initial velocity is assigned to them, as well as channels of communication between the particles. These particles then move in the response space, and the results are calculated based on a "goal function" after each time. Over time, particles accelerate toward particles that have a higher competency standard and are in the same communication group. Each agent knows its best value so far (pbest) and its XY position. Moreover, each agent knows the best value in the group (gbest) among pbest s. By initializing the *swarm* from the solution space, the velocity and position of all particles are randomly set to within determined ranges. At any iteration the velocities of all particles are updated. The velocity of each agent can be modified by the following equation: [12, 13]

$$v_{i+1} = v_i + c_1 R_1 (p_{i,best} - x_i) + c_2 R_2 (g_{i,best} - x_i) \tag{5}$$

Where $x_i$ and $v_i$ are the position and velocity of particles current position and velocity. R1 and R2 are the uniformly distributed random numbers in the range [0–1], which introduce the stochastic component. $c_1$ and $c_2$ are factors controlling the related weighting of corresponding terms. Using the above equation, a certain velocity, which gradually gets close to pbest and gbest that has been found by all the particles in the swarm, can be calculated. The position update equation is given by:

$$p_i new = p_i + v_i \tag{6}$$

After updating, $p_i$ should be checked and limited to the allowed range. Afterwards, when condition is met Update $p_{i,best}$ and $g_{i,best}$ as following:

$$\begin{aligned} p_{i,best} &= p_i & \text{if } f(p_i) > f(p_{i,best}) \\ g_{i,best} &= g_i & \text{if } f(g_i) > f(g_{i,best}) \end{aligned} \tag{7}$$

where *f(x)* is the objective function to be optimized. When stop conditions were met, algorithm reports the values of $g_{i,best}$ and $f(g_{i,best})$ as its solution. The algorithm goes on until a satisfactory solution is found or the predefined number of iterations is met.

### *3.3. Ant Colony Optimization (ACO)*

ACO is one of the most recent techniques for approximate optimization [14-17]. In this method (ACo), artificial ants by moving on the problem diagram and by leaving marks on the diagram, like real ants that leave marks in their path, make the next artificial ants can provide better solutions to the problem. Also in this method, the best path in a diagram can be found by computational-numerical problems based on probability science. While an isolated ant moves practically at random, exploration, an ant encountering a previously laid trail can detect it and decide with high probability to follow nit, exploitation, and consequently reinforces the trail with its own pheromone.

### 4. Results and Discussions

In this section, to investigate the accuracy and truth of the proposed method some conditions were considered and simulated. The simulation results of three optimization algorithms and conventional constant weighting factor control are presented. The simulations are performed in MATLAB/SIMULINK environment under the following conditions:

- Variation of DC-DC converter output load (at +5 V output)
- Variation of DC-DC converter output load (at +15 V output)
- Variation of DC-DC converter input voltage

It should be noted that the fuzzy logic controller has been used for all approaches including constant weighting factors (which were calculated by conventional mathematical method) and all three proposed methods at three mentioned conditions. Also, for the same system, a PID controller is designed. The total error is subsequently passed through a PID controller to minimize the steady-state error as well as a fuzzy logic controller in section 4.5 when the PID coefficients are determined for the studied DC-DC converter by using trial and error. Also in this section, the performances of three used algorithms are compared with each other.

### *4.1.   Variation Of DC-DC Converter Output Load (At +5 V Output) From 100% to 50%*

In this case, the load on +5 v output of the converter changes from 100% to 50% at t=10 ms when the output voltages are in steady state. Fig.5 (a) shows three output voltages when ICA is used as a weighting factors estimator. Figs.5 (b) and (c) show these output voltages for ACO and PSO weighting factors estimator, respectively. Accordingly, Fig.5 (d) shows these voltages for constant weighting factor are used for voltage mode weighting factor control. These results show that the use of an evolutionary algorithm can improve the percent of cross-regulation significantly.

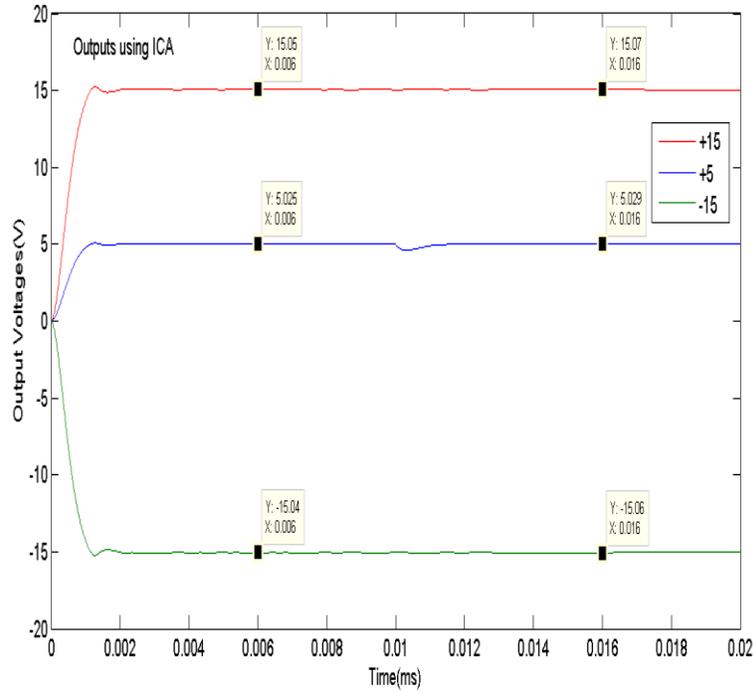

(a)

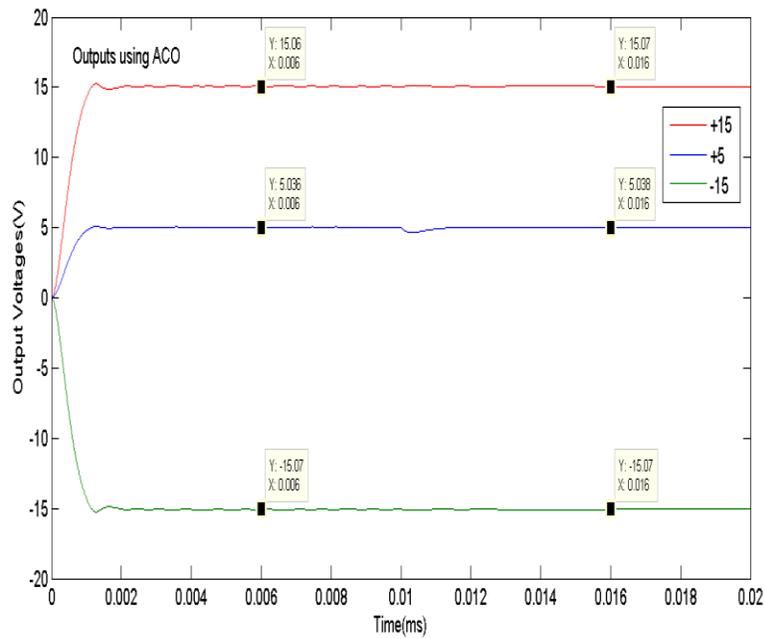

(b)

**Fig.5.** Output voltages while load on +5 V output changes from 100% to 50% with a) ICA estimator of weighting factors, b) ACO estimator of weighting factors.

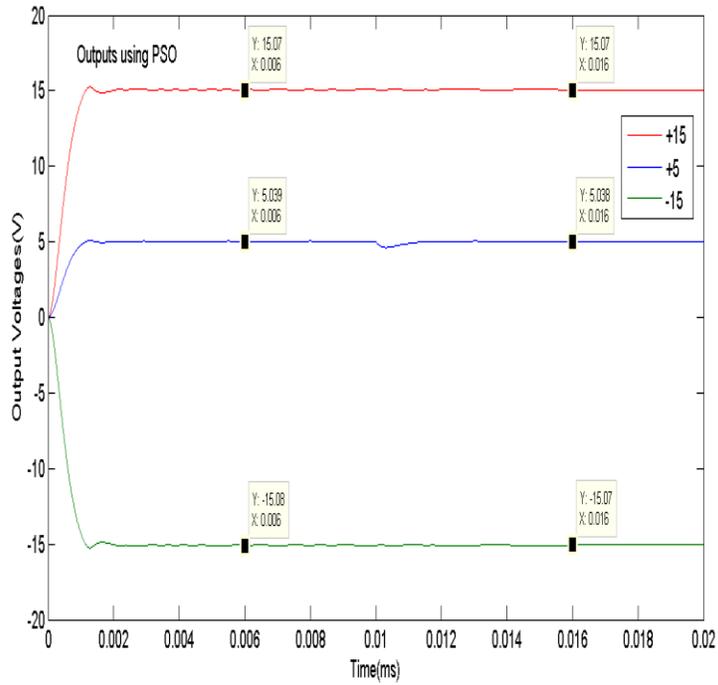

(c)

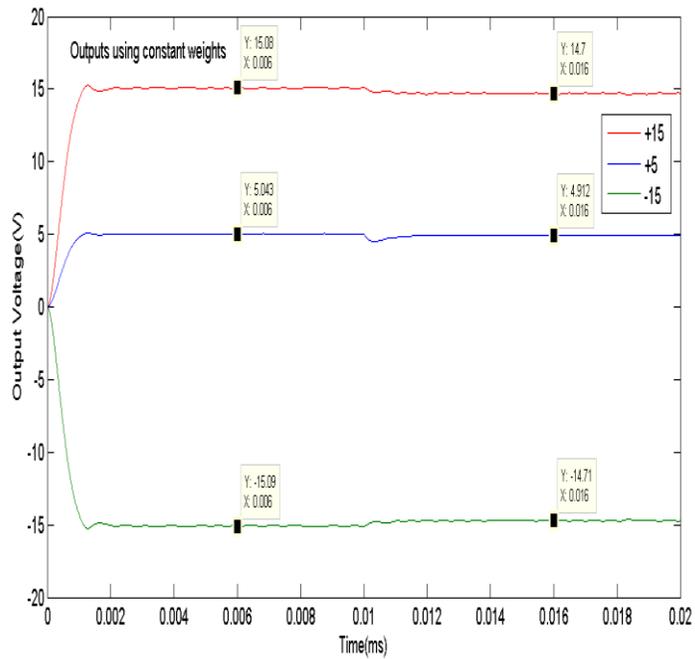

(d)

**Fig.5.** Output voltages while load on +5 V output changes from 100% to 50% with c) PSO estimator of weighting factors, d) constant weighting factors (Continue).

Table 2 shows the regulation of the presented methods before and after load changing. In a constant weighting factor method, the weighting factors are determined by trial and error at the system condition before load changing (as t=6 ms) for optimum regulation. Thus, the regulation can be good in this condition but after changing of conditions, regulation may be reduced whereas three other methods have lower regulation. Also because of its higher accuracy and speed, ICA based weighting factor estimator is more effective in comparison with ACO and PSO based weighting factor estimator.

Table 2. Output voltage regulation while load changes at +5 v output from 100% TO 50%.

| Method | Regulation (%) at t=6 ms | | | Regulation (%) at t=16 ms | | |
|---|---|---|---|---|---|---|
| | +15V | +5V | -15V | +15V | +5V | -15V |
| ICA | 0.33 | 0.50 | 0.26 | 0.46 | 0.58 | 0.40 |
| ACO | 0.40 | 0.72 | 0.46 | 0.46 | 0.76 | 0.46 |
| PSO | 0.46 | 0.78 | 0.53 | 0.46 | 0.76 | 0.46 |
| Constant Weighting Factor | 0.53 | 0.86 | 0.60 | 2 | 1.76 | 1.93 |

### 4.2. *Variation Of DC-DC Converter Output Load (At +15 V Output) from 100% To 120%*

In the third case, the load on +15 v output of the converter changes from 100% to 120% at t=10 ms when the output voltages are in steady state. Figs.6 (a), (b), (c) and (d) show the three output voltages for ICA, ACO and PSO based estimator of weighting factor as well as constant weighting factor method, respectively. Table 3 shows the output voltage regulation. This result also shows that the proposed method can improve the cross-regulation in all three outputs. Results show that the ICA can be more accurate than two other algorithms.

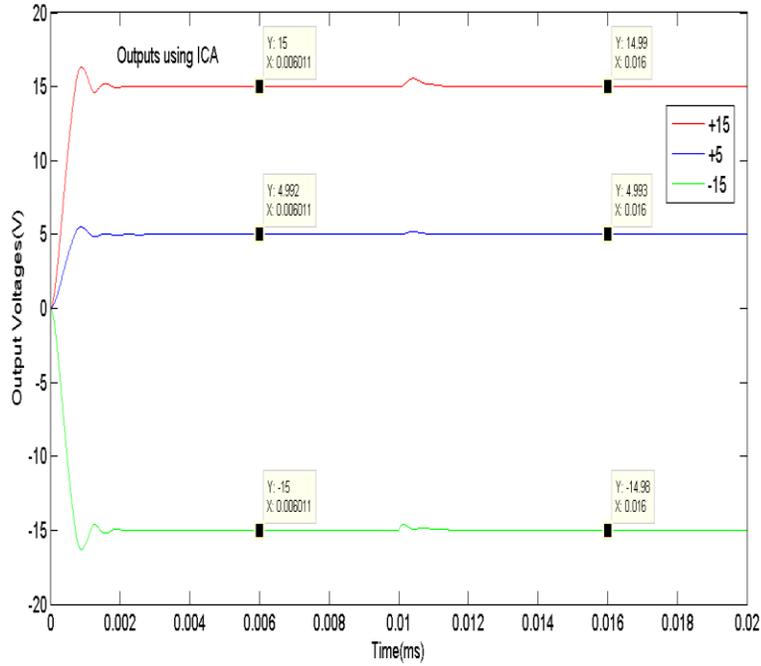

(a)

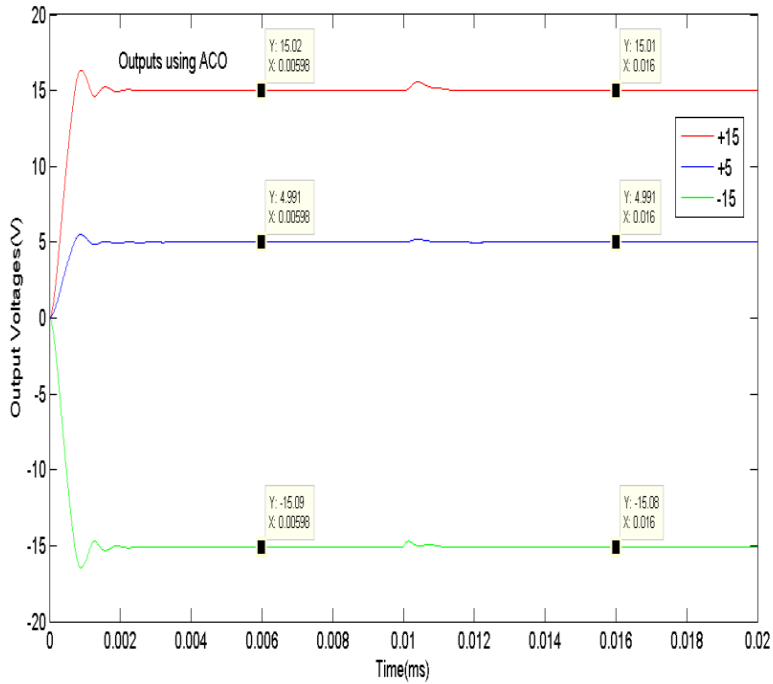

(b)

**Fig.6.** Output voltages while load on +15 V output changes from 100% to 120% with a) ICA estimator of weighting factors, b) ACO estimator of weighting factors.

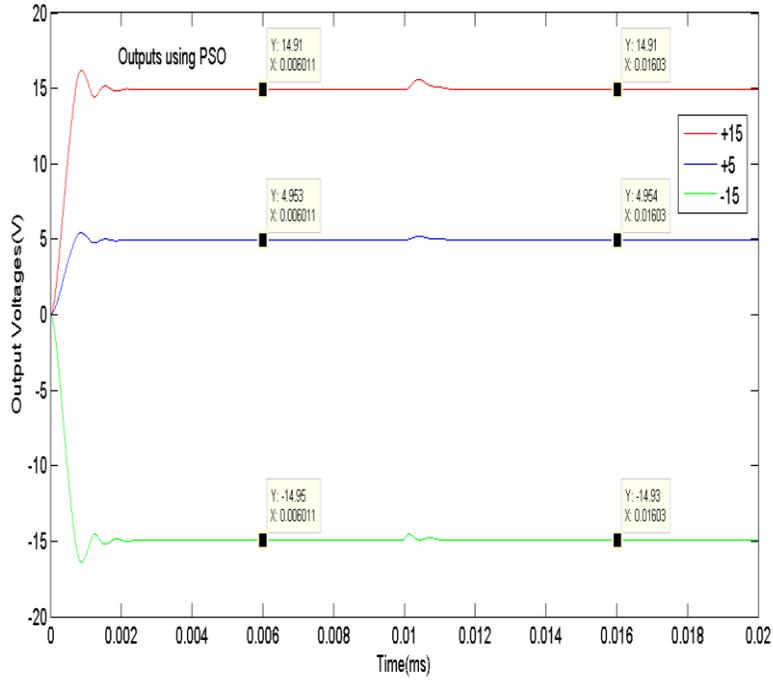

(c)

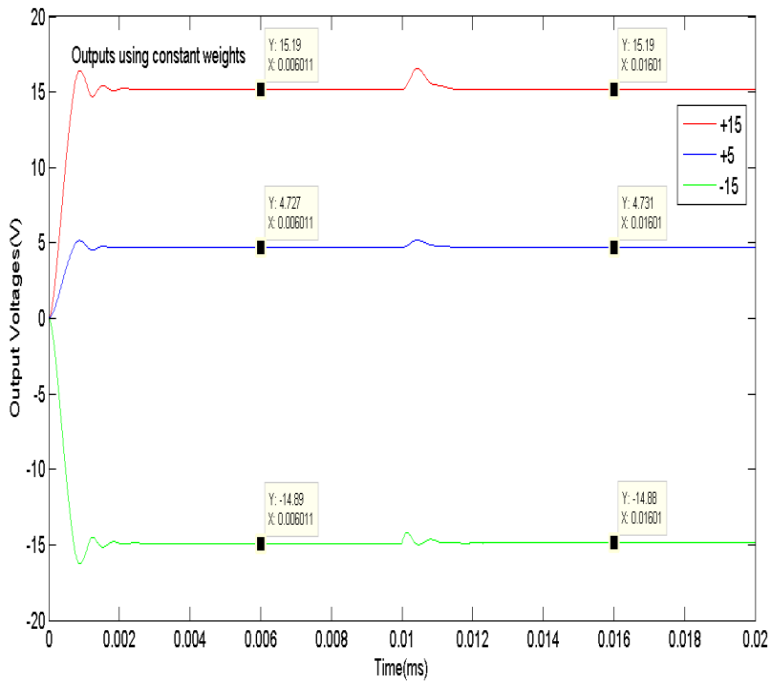

(d)

**Fig.6.** Output voltages while load on +15 V output changes from 100% to 120% with c) PSO estimator of weighting factors, d) constant weighting factors (Continue).

Table 3. Output voltage regulation while load changes at +15 v output from 100% to 200%.

| Method | Regulation (%) at t=6 ms | | | Regulation (%) at t=16 ms | | |
|---|---|---|---|---|---|---|
| | +15V | +5V | -15V | +15V | +5V | -15V |
| ICA | 0.00 | 0.05 | 0.00 | 0.06 | 0.14 | 0.13 |
| ACO | 0.13 | 0.18 | 0.60 | 0.06 | 0.18 | 0.53 |
| PSO | 0.60 | 0.94 | 0.33 | 0.60 | 0.92 | 0.46 |
| Constant Weighting Factor | 1.26 | 5.46 | 0.73 | 1.26 | 5.38 | 0.80 |

### *4.3.  DC-DC Converter Input Voltage Changing From 30 V to 35 V*

In this case, input DC voltage of DC-DC converter changes from 30V to 35V. Figs.7 (a), (b), (c) and (d) show the three output voltages for ICA, ACO and PSO based estimator of weighting factor as well as constant weighting factor method, respectively. Results of this case show that the evolutionary algorithms are very effective for improving cross-regulation in all outputs. Results after changing at t=16 ms show that the constant weighting factor method is not a suitable method for all conditions.  The results of regulation for comparison of all methods are presented in Table 4.

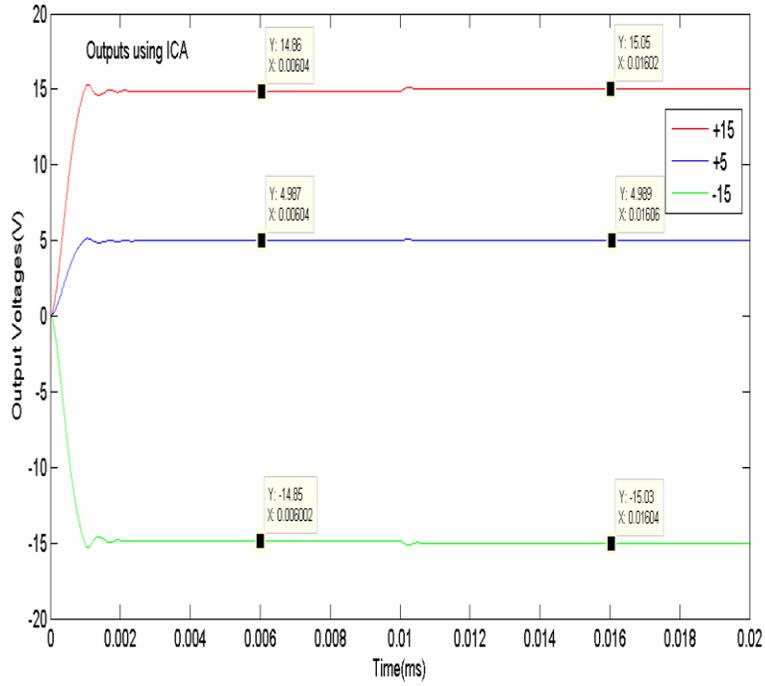

(a)

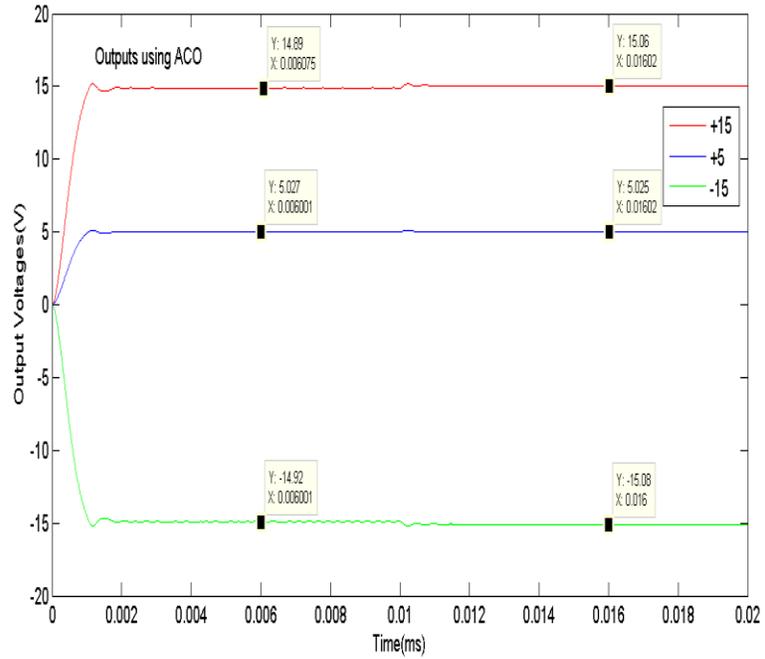

(b)

**Fig.7.** Output voltages while input voltage changes from 30 V to 35 V with a) ICA estimator of weighting factors, b) ACO estimator of weighting factors.

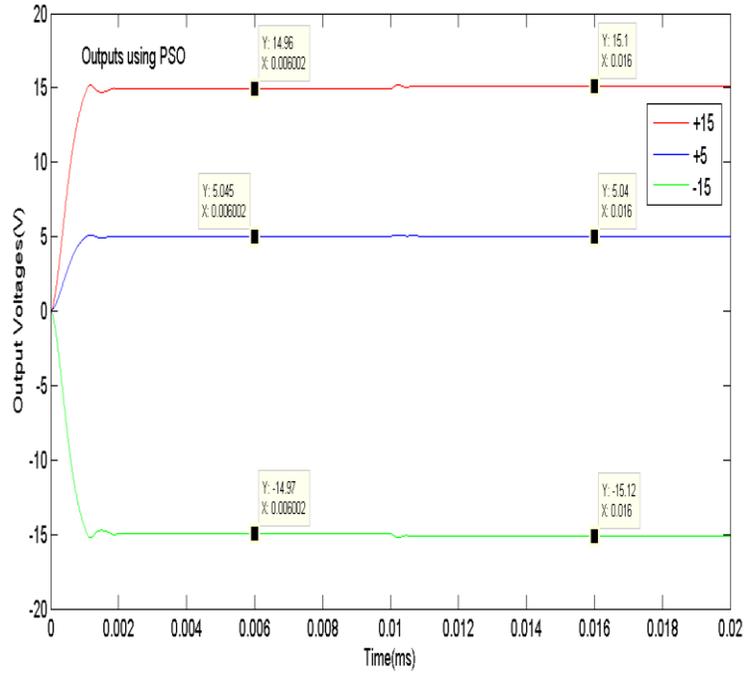

(c)

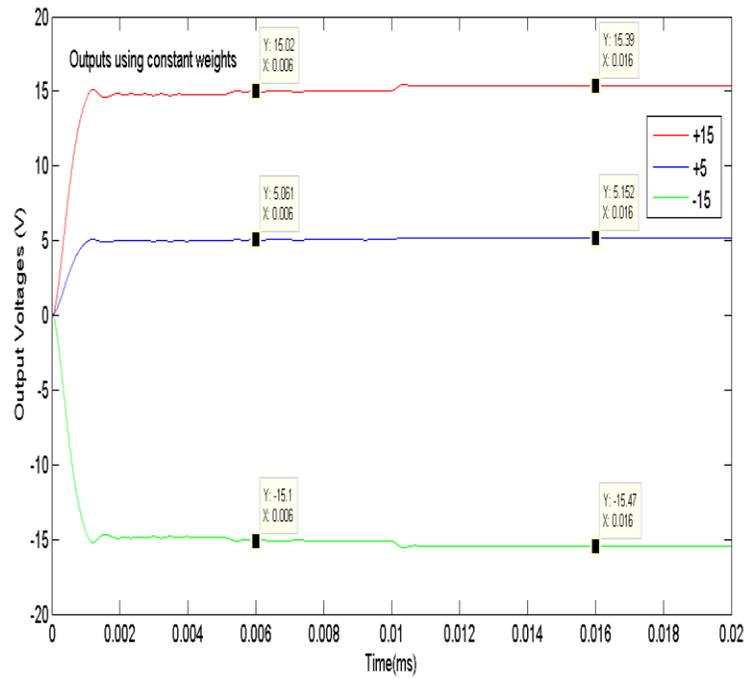

(d)

**Fig.7.** Output voltages while input voltage changes from 30 V to 35 V with c) PSO estimator of weighting factors, d) constant weighting factors (Continue).

Table 4. Output voltage regulation while input voltage changes from 30v to 35v.

| Method | Regulation (%) at t=6 ms | | | Regulation (%) at t=16 ms | | |
|---|---|---|---|---|---|---|
| | +15V | +5V | -15V | +15V | +5V | -15V |
| ICA | 0.93 | 0.26 | 1.00 | 0.33 | 0.22 | 0.20 |
| ACO | 0.73 | 0.54 | 0.53 | 0.40 | 0.52 | 0.53 |
| PSO | 0.26 | 0.90 | 0.20 | 0.66 | 0.80 | 0.80 |
| Constant Weighting Factor | 0.13 | 1.22 | 0.66 | 2.26 | 3.04 | 3.13 |

## *4.4.     Dynamic Response Comparison*

The +5 V output voltage results of mathematical based weighting factors estimator, together with a set of optimal weighting factor that is obtained from one of the best used evolutionary algorithm (ICA), are presented in Fig.8 when both fuzzy logic controller and PID methods are examined their performances have been investigated via simulations. The time response parameters percent overshoot (%), settling time (ms) and percent steady-state error (%) for PID controller and fuzzy logic controller when ICA and conventional mathematical method has been used for both as weight factor estimator, are presented in Table 5. As shown in this figure the choice of appropriate weighting factors, in addition to improving the voltage regulation can positively affect the dynamic behavior of the system. Results show that when the optimal weighting factors are used, the overshoot value and settling time have reduced significantly.

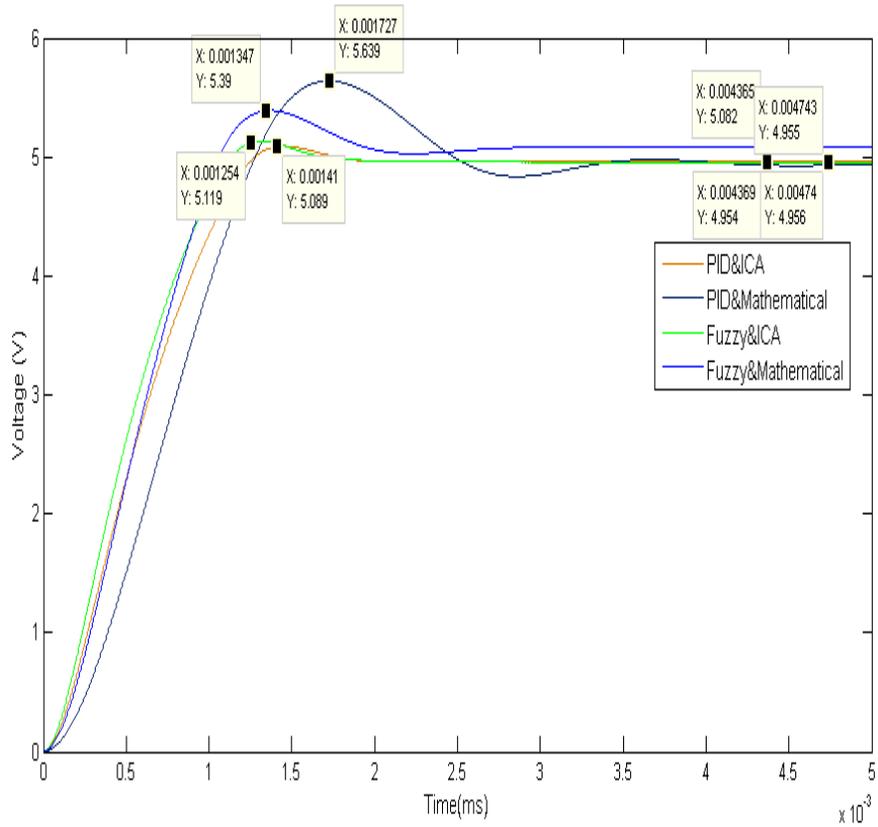

**Fig.8.** The +5 V output voltage results for Fuzzy ICA, PID ICA, Fuzzy mathematical and PID mathematical methods.

Table 5. Time response parameters.

| Method \ Item | Over shoot (%) | Settling time (ms) | Steady state error (%) |
|---|---|---|---|
| **Fuzzy & ICA** | 2.38 | 1.9 | 0.88 |
| **PID & ICA** | 1.78 | 1.9 | 0.90 |
| **Fuzzy Mathematical** | 7.8 | 2.5 | 1.64 |
| **PID & Mathematical** | 12.78 | 3.5 | 0.92 |

## 4.5. Comparison of the Presented Algorithms Performances

In this section, the performance of three used algorithms is compared with each other. Table 6 shows the performances of the algorithms at different conditions. Total regulation in this table is the sum of the regulation at three output voltages. Based on performed simulations, it is shown that the cross-regulation has been better done via ICA respecting the PSO and ACO. Besides, it has been demonstrated that although almost all of the optimization methods are able to improve the cross-regulation properly, the ICA out-performs the others in terms of the total regulation, execution time and convergence rate.

Algorithms are coded in MATLAB and are a helping factor to guide the exploration and to increase the run with a Pentium 2.4GHz computer. In all the examples, control in the exploitation of the population is equal to 50. And the maximum iteration is 50 for all algorithms. ICA appears to offer good applicability in this work and further applications should be explored.

Table 6. Algorithms performances at different conditions.

| Conditions | Convergence Iteration | | | Total regulation % at t=16ms | | |
|---|---|---|---|---|---|---|
| | ICA | ACO | PSO | ICA | ACO | PSO |
| **Load variation from 100% to 50%** | 75 | 85 | 100 | 1.44 | 1.68 | 1.68 |
| **Voltage variation from 30V to 35V** | 75 | 80 | 110 | 0.75 | 1.45 | 2.26 |
| **Load variation from 100% to 120%** | 70 | 85 | 110 | 0.33 | 0.77 | 1.98 |

## 5. Conclusion

A design procedure is established for the estimation of weighting factors in weighted voltage mode control of multiple outputs forward DC-DC converter. The innovation of this paper is the use of evolutionary algorithms for determining weighting factors in weighted voltage mode control approach to optimize regulation of multiple outputs DC-DC converter. Accordingly, Imperialist Competitive Algorithms, Particle Swarm Optimization, and Ant Colony Optimization are utilized to find the optimal weighting factor for voltage mode control of typical multiple outputs forward DC-DC converter. Also, the fuzzy logic controller has been utilized to improve the dynamic response of the converter. In order to compare the performance of different methods, results are presented in converter input and output variations. This approach does not require any parameterization or structural formalization of the modeling uncertainty.

Results show that the proposed method can improve cross regulation and dynamic characteristics of the outputs significantly in comparison with the constant weighting factor method. The ICA based weighting factor estimator has higher speed and accuracy in comparison with two other presented evolutionary algorithms (ACO and PSO) and so is more effective in comparison with them. Results of the simulation show that if the weighting factors are properly designed, the regulation and dynamic characteristics of the sensed outputs can be significantly improved. The proposed method can be used for other types of multiple output DC-DC converters.